Identity, Kinship, and the Evolution of Cooperation

Burton Voorhees[1*], Dwight Read[2], and Liane Gabora[3]

[1] Center for Science, Athabasca University, 1 University Drive, Athabasca, AB, T9S 3A3; burt@athabascau.ca, http://science.athabascau.ca/staff-pages/burtv/

[2] Department of Anthropology, UCLA, 375 Portola Plaza, 341 Haines Hall, Box 951553, Los Angeles, CA 90095; dread@anthro.ucla.edu, http://www.anthro.ucla.edu/faculty/dwight-read

[3] Department of Psychology, University of British Columbia, Okanagan Campus, Kelowna, BC Canada V1V 1V7; liane.gabora@ubc.ca, https://people.ok.ubc.ca/lgabora/

* Please address correspondence to Burton Voorhees, burt@athabascau.ca

**Abstract**

Extensive cooperation among biologically unrelated individuals is uniquely human and much current research attempts to explain this fact. We draw upon social, cultural, and psychological aspects of human uniqueness to present an integrated theory of human cooperation that explains aspects of human cooperation that are problematic for other theories (e.g., defector invasion avoidance, preferential assortment to exclude free riders, and the second order free rider problem). We propose that the evolution of human cooperative behavior required (1) a capacity for self-sustained, self-referential thought manifested as an integrated worldview, including a sense of identity and point of view, and (2) the cultural formation of kinship-based social organizational systems within which social identities can be established and transmitted through enculturation. Human cooperative behavior arose, we argue, through the acquisition of a culturally grounded social identity that included the expectation of cooperation among kin. This identity is linked to basic survival instincts by emotions that are mentally experienced as culture-laden feelings. As a consequence, individuals are motivated to cooperate with those perceived culturally as kin, while deviations from expected social behavior are experienced as threatening to one's social identity, leading to punishment of those seen as violating cultural expectations regarding socially proper behavior.

## Introduction

Humans cooperate to an extent that defies biological expectations. As noted by Darwin (1871:179), human cooperation is evidenced by traits that include patriotism, fidelity, obedience, courage, sympathy, and willingness to sacrifice for the common good, which we will refer to collectively as *ultrasociality*. How did individuals acquire these traits given that they may impose strong, even fatal burdens on individual fitness? And can such traits persist in a population without succumbing to free riders and cheats?

Yet, ultrasociality is a universal characteristic of human societies (Turchin 2013). Even in groups of close biological kin, humans are more cooperative than would be expected if genetic relatedness and/or direct reciprocity were the only factors involved (Mathew *et al* 2013). Although humans may act in spectacularly uncooperative ways, and non-human animals sometimes appear to act cooperatively (Packer 1977; Watts 2002; Clutton-Brock 2009; Silk, *et al* 2013), human cooperation is unique in its degree and scope. A particularly significant aspect of human cooperation is its flexibility. Humans not only cooperate in large groups of biologically unrelated individuals, there is a great deal of flexibility as cooperative behavior is adjusted to adapt to new circumstances.

There have been extensive efforts to resolve the paradox of human cooperation by explaining it in terms of individual traits arising through culture-gene coevolution (Richerson and Boyd 2005; Richerson, *et al* 2010; Bowles and Gintis 2011; Mesoudi 2011; Richerson and Christensen 2013; Richerson, *et al* 2016). Culture-gene coevolution is modeled in terms of inclusive fitness (Taylor, *et al* 2007; West and Gardner 2013;

Birch and Okasha 2014; Bourke; 2014; Okasha, *et al* 2014) and/or multilevel selection (Wilson and Sober 1994; Okasha 2005; O'Gorman, *et al* 2008; Gardner 2015). Despite extensive debate among advocates of these two approaches (e.g., West, *et al* 2007; Kramer and Meunier 2016), they have been shown to be predictively (but not causally) equivalent (Okasha 2016).

In biological evolution the mechanisms of variation and selection operate on individual phenotypes, resulting in differential trait distributions across a population. When an analogue of this model is applied to cultural evolution, it has been found that selection acting on group level traits becomes a significant factor in addition to selection acting on individual traits. Current theorizing, based on culture-gene coevolution, has focused on reconciling the consequences of these two different levels of selection. In culture-gene coevolution theories, groups compete for resources, or for survival in a harsh environment, and group selection favors groups in which individuals coordinate their cooperative activities. This sets up a tension between within-group selection, which tends to favor selfish individual traits, and *cultural* group selection, in which the group level traits acted on are cultural in nature so that selection favors cultural traits arising from the expression of more altruistic or prosocial individual traits[1].

The critical distinction between humans and other animals in culture-gene theories is the capacity for cumulative cultural transmission (Gabora 1998; Boyd 2018). Cultural traits are said to be selected for based on their utility in satisfying necessities of individual and group survival. Culture, in turn, produces a social and material environment such that group-beneficial cultural traits will tend to be tuned to human psychological and behavioral tendencies. As expressed by Henrich and Boyd (2016), "Intergroup competition will favor those group-beneficial cultural traits… that most effectively infiltrate and exploit aspects of our evolved psychology." This is said to lead to genetic adaptation for individual traits such as imitative behavior, prosocial preferences, and norm following, all of which, it is argued, contribute to cultural adaptation and transmission (Boyd and Richerson, 1985; Boehm 1999, 2012; Richerson and Boyd, 2005; Boyd, *et al* 2005; Chudek and Henrich 2011; Sterelny 2012; Bowles and Gintis, 2011; Heyes 2012).

These theories, however, fail to provide an integrated explanation for several key aspects of human cooperation, which this paper aims to address. These include: (1) how the selective balance between cultural group selection and within-group selection on individuals is shifted to favor cooperation enhancing behavior by individuals within a cultural group; (2) how this is maintained against invasion by defectors; and, (3) what stabilizing factors might be in operation. Explanations offered by existing culture-gene theories appeal to cultural transmission, positive assortment of cooperators, punishment of defectors, and appeal to the social emotions (e.g., guilt and shame). However, they neither offer an explanation of how these factors actually promote stable cultural transmission of cooperative behavior nor of how stable transmission relates to unique elements of human biology and psychology. The culture-gene approach has been characterized as the "puzzle paradigm" and may be incapable of providing these explanations (Cowden, *et al* 2017). Dual inheritance theories focus on trait evolution whereas it is necessary to address the "co-evolution of human cognition, societal organization and society's engagement with the environment (Read and van der Leeuw 2015)" (Sterelny, 2012; van der Leeuw 2018:2).

Other than the initial assumption of cumulative cultural transmission, culture-gene theories say little about human uniqueness. Analysis is carried out using methods that could apply to any species. Appealing to cultural traits that exploit an evolved human psychology does not, in and of itself, address how this psychology actually produces cooperative behavior. Nor does it say anything about the relationship between cultural traits, individual psychology, and human biology. It cannot explain why social emotions exert such a strong influence on individuals, how these emotions are influenced by culture, or how culture coopts human biological instincts so as to produce cultural, and in particular cooperative behavior.

**Outline of Proposed Theory**

To explain how individuals conceive of, and understand, both their behavior and that of others requires a shift in perspective to thinking in terms of top down culturally prescribed patterns of coordination and cooperation. We explain cooperation not in terms of traits at the individual level that yield cooperation as an emergent trait at the population level, but in terms of *cultural idea systems* that constitute the "organizational or conceptual universe of a community and the principles that govern it, that make it coherent and thereby make it objective for its adherents" (Leaf and Read 2012: 48). Cultural idea systems frame the structure and organization of social interaction as structured patterns of affectively salient prescriptions, prohibitions, and interpretations, as well as stories, schemas, myths, narratives, and understandings that become incorporated as elements of the worldviews of group members[2] (Read 2003; Leaf and Read 2012; Gabora, 2004).

Individual worldviews are self-sustaining internalized webs of knowledge, beliefs, attitudes, ideas, affections, and expectations (Gabora 2004) arising through interactions between personal experience and the experience of being enculturated into a system of cultural ideas that informs, guides, and provides a common means for evaluating behaviour (Leaf and Read 2012). From this conceptual foundation of worldviews and cultural idea systems, transmitted through enculturation, we can develop explanations relating culture, human psychology, and human biology in ways that show the interconnections between evolved human psychology and culturally mediated human behavior.

We propose that the foundational factor for human uniqueness is the evolution of brains capable of supporting *reflective self-consciousness*. We distinguish between self-consciousness and reflective self-consciousness in a way similar to Edelman's (2004) distinction between primary consciousness and higher-order consciousness. Beyond a simple awareness of sensations, by *reflective self-consciousness* we mean being aware of oneself as the individual self that is having those sensations, together with the capacity to reflect on that self both historically and as projected into the future, including perspective taking, analogy making, and the capacity for self-triggered recall (Gabora and Steel 2017; Gabora and Smith *in press*). It is this that enabled humans to assimilate elements of experience into a coherent, culture-laden understanding of the world—a *worldview*—that included a consciously recognized social identity defined by group membership and social roles. With this, cultural systems of social relations, in particular kinship systems, developed as determining factors in human behavior (Read 2012a; Moffett 2013). Kinship systems are paradigmatic exemplars of what we mean by cultural idea systems.

In the manner that group-level traits are understood in much of the culture-gene literature, functionality for the group emerges from the individual level as group members "function cohesively to maximize fitness (reproductive success) at the higher level" (O'Gorman *et al* 2008). In this bottom-up framework, culture arises through individual phenotypic behavior and sets the environment within which individual traits evolve. Synthesis of these traits over the group yields emergent group level traits. In this framework, cooperative behavior is grounded in individual phenotypic traits and so it supposedly can be treated with the methods of evolutionary population biology; for example, through appropriate partitioning of terms in the Price equation (Kramer and Meunier 2016; Okasha 2016).

In contrast, cultural idea systems are not emergent from individual traits. Rather, they are top-down schemas for how individuals are to understand events and act (Read 2003; Leaf and Read 2012). Therefore, though a cultural idea system manifests itself in a group-level phenotypic manner, this differs qualitatively from the way others (e.g., Chudek and Henrich 2011; Smaldino 2014; Richerson *et al* 2016) consider group-level traits to be emergent from individual traits.

Cultural idea systems provide group members with synchronized modes of attention, interpretation, and emotional attunement that allow group syntonization[3] for cooperative action. These systems permeate a culture: individuals are enculturated from birth with the behavioral, institutional, and normative instantiations of these systems. Through this enculturation, cultural idea systems exert a formative influence on individual worldviews. Enculturated individuals experience the behavior, interpretations, institutions, and norms involved in the social expression of cultural ideas as parts of the natural and social order. In this way, cultural idea systems serve the essential role of enabling mutual understanding among members of the social collective. As such, they can be sources of functionality at the individual level because individuals gain benefits by conforming to and contributing to group level cultural schemas and prescribed roles (Read 2012a).

Cultural idea systems and individual worldviews are intimately connected. Worldviews are the medium through which cultural idea systems are transmitted across generations, anchored by material and social forms (symbols, monuments, institutions, rituals) that gain meaning only as they are understood by individuals whose worldviews have been enculturated within the corresponding cultural idea systems as children are taught the cognitive and affective meaning of external cultural manifestations and social behavior (e.g., Quinn 2006).

Focusing on cultural idea systems as vehicles of individual understanding and cooperative behavior requires demonstration of strong connections among culture, individual psychology, and human biology. We find such connections by distinguishing between *emotions*, defined as physiological reactions to salient cues, and the culture-laden mental *feelings* that accompany these physiological reactions (Damasio 2010; LeDoux 2012; Barrett 2017). Socially appropriate responses to contextually evoked emotional reactions are mediated by cultural idea systems that provide context dependent interpretations of, and proper behavioral expressions for, the associated feelings[4] (Voorhees *et al* 2018).

The emotion/feeling connection is two-way. External cues can elicit culturally formulated feelings that trigger an emotional response (Damasio 2010). Thus, threats to a culturally based social identity, or to the group identity within which social identity is embedded, may evoke physiological reactions as if the threat were directed at the biological organism. This can lead to a defensive impulse to punish those perceived as going against cultural ideals or norms shared by group members (Cikara, *et al* 2011; Cikara and Van Bavel, 2014). As a result, there is no second-order free rider problem in our theory. That is, a second-order free rider problem exists only if there is uncertainty as to whether group members who have agreed to punish defectors will actually fulfill this commitment should the occasion arise. In our theory, the impulse to punish free riders and defectors arises automatically as a defensive response to a perceived threat to one's social identity. This establishes both internal and external controls on cheating. Cheaters risk punishment by threatening the social identity of their cultural kin even as they distance themselves from the social group upon which survival depends[5].

When individuals become self-identified as members of a social group, this membership is incorporated into their social identity and behavioral expectations regarding other group members are seen as cultural obligations. The relevant difference between an expectation and an obligation is that the former, if unmet, may result in an individual frustration/anger/rage reaction while the latter, if violated, invites group-supported sanctions against the violator.

In sum, because current theories of human cooperation do not address how minds and societies function as integrated wholes, they cannot account for several aspects of human cooperation that we address here. We posit that individuals gain social identity through group membership and role performance; this identity is conditioned by cultural idea systems that structure individual worldviews and mandate behavioral expectations and obligations, including cooperation with cultural kin. Moreover, we propose that the link between biological emotions and culturally laden feelings results in cooption of defensive reactions based in survival related neural circuits for the defense of both individual social identity and the groups from which this identity derives.

## Identity, Point-of-View, and Kinship

We now consider the concept of individual identity and its expression as a point of view, centered in an individual worldview. We argue that cooperative behavior requires not only the recognition that other group members also have points of view that may differ from our own, but also a means of bringing disparate points of view into alignment. While incidents of signaled group identification and joint intentionality (Moffett 2013;

Tomasello 2014) are precursors of collective cooperation, we suggest that it was the advent of cultural kinship systems among our Paleolithic ancestors that provided the keystone for stabilizing individual points of view into a collective cultural understanding (see Read 2012a).

Some non-human primates appear to be self-aware according to the mirror test (Gallup Jr. 1970; Povinelli et al. 2003). Even without this recognition, however, primate awareness can be exquisitely sensitive to social nuances, yet remain unreflective. Seyfarth and Cheney, for example, (2000:902) suggest that: "Although monkeys may behave in ways that accurately place themselves within a social network, they are unaware of the knowledge that allows them to do so: *they do not know what they know, cannot reflect on what they know, and cannot become the object of their own attention*." (emphasis added)

In contrast, humans do not merely pursue their own benefit; they are aware of their identity as individuals, and as members of a social group. Humans are *persons*, in a way that other apes are not. A human becomes a *person* not through biological birth, but by developing the capacity for self-triggered thought and an integrated worldview with a point of view that becomes aligned with the point of view of kin, which in turn involves being integrated into a cultural group through group recognition of their personhood. Humans recognize themselves both as individuals and as members of a social collective, and they are recognized as such by other members of this collective. This means humans possess something more than what is commonly called theory of mind. *Theory of mind* refers to the capacity to recognize that others have mental states (Tomasello 2014). There is a critical difference, however, between recognizing that others have mental states, and understanding that not only is there another creature with mental states, but this other creature is like oneself (e.g., as a member of a conceptualized group), thus is simultaneously different (not–self), yet similar (a creature of the same kind). This requires the cognitive ability to abstract from a collection of similar objects or entities and grasp a categorical concept binding that collection into an abstracted unity (Read 2012b; Gabora and Smith *in press*). It is this that allows members of a human foraging band to conceptualize "we" in reference to the band as a whole, something that appears to be outside the purview of chimpanzees as suggested by the fact that chimpanzee communities do not act as a social unit[6] (Whiten and Erdal 2012).

A more complex situation arises with awareness that others not only have mental states, but also a point of view. It might seem that attribution of mental states automatically includes attribution of a point of view, but this need not be the case. To imagine that another individual has a point of view requires being able to abstract from ones' own mental states to the recognition of a self with the capacity to reflect upon those states, then attributing this capacity to others. The other becomes an individual rather than just a locus of mental states. We refer to the recognition that others not only have mental states, but also points of view as *theory-of-person.* When Tomasello (2014) proposes "joint intentionality" as a prime factor leading to human cooperative behavior and relates this to theory of mind, he is appealing, in our terms, to theory-of-person.

It can be argued that acting so as to deceive another by inducing a false belief (behavior found in other primates) indicates awareness of the other as having a point of view that can be misdirected, but it is more parsimonious to see this as simply acting so as to produce a mental state in another in order to achieve an immediate selfish purpose. Theory-of-mind without theory-of-person invites deception; it leads to the impulse to produce a desired result by manipulating the mental states of others. No more substantial recognition of the other is required. Theory of person leaves room for cooperation—it leads to the question: "How can *I* act in *our* mutual best interest when *our* points of view are in alignment?" Because cultural idea systems are distributed among group members, the worldviews of culture-bearers share a common understanding of the meaning of behavior arising through those idea systems, facilitating alignment in points of view (Read 2012b). Hence, three pre-conditions are necessary for our theory of cooperation: (1) the coalescing within individuals of experiences into an integrated worldview with a point-of-view, (2) theory-of-person and (3) a means of aligning different points-of-view. These conditions require cognitive capacities unique to late members of the genus *Homo* (Read 2008; Read and van der Leeuw 2008).

The best candidate for how disparate points of view could have been brought into alignment among our Paleolithic ancestors in a stable and transmissible way is through the formation of a system of culturally expressed kinship relations as the basis for social organization, group identity, and group boundaries (Read 2012a,b). Kinship provides a system of conceptual relations interconnecting individuals in a manner that enables the perspective of one individual to be transformed into that of another (Read 2003). It gives answers to the individual question "Who am I?" and the social question, "Who are We?" by transforming biological birth into a position within a cultural kinship system (Sahlins 2013). This cannot be reduced to biological relations since the cultural importance of birth is not that it produces a member of the species *Homo sapiens*, but that it yields a person with a social identity, distinguishing an infant as a *person,* as "one of us." As a child develops to maturity within a cultural context, being "one of us," together with the concomitant behavioral, ideational, and affective conditions and constraints that this implies, becomes an intrinsic aspect of the child's identity[7].

**Cultural Idea Systems, Identity, and Worldview**

The variety of world cultures is large, but there are fundamental elements in all cultures—kinship, religion, hierarchical striving, and political systems of social organization, to mention a few. These commonalities arise because culture provides templates, schemas, and scripts, as well as stories and myths, for socially productive transformations of universal human instinctual drives (Goddard and Wierzbicka 2004; Matsumoto 2007; Tsai 2007; Trommsdorff and Cole 2011; De Leersnyder, *et al* 2013). From birth onward, processes of enculturation embed these patterns in individual worldviews in the form of prescribed and proscribed behavior, beliefs, and attitudes that come to be experienced as natural, obvious and unquestionable (Spradley and Mann 1975; Weissner 1998).

In the processes by which infants individualize themselves by formulating personal ideas and perspectives (Barnett 1953), cultural idea systems provide organizing patterns that give internal structures of belief and interpretation together with external channels for socially acceptable behavioral expression of basic concerns, wants, desires, and needs.

Core elements of a worldview are affectively tagged constructs of personal identity, group identity, and individual and group purpose (Markus and Kitayama 1991, 2010; Mantovani 2000; Narvaez 2013). Human children are taught not only the necessities of immediate survival, but also how to think symbolically about the world. They are taught the rites, practices, attitudes, and beliefs that sustain both personal and cultural identities. They are taught what behavior deserves respect, what is decent, and what are worthwhile achievements (Mantovani 2000; Handerwerker 2015). Culture forms the warp and weft of their thought, upon which the tapestry of identity is woven.

What is the ontology of the cultural ideas systems that structure individual worldviews in ways that provide for mutual understanding within the cultural group? The answer, in many cases, is recursive—cultural idea systems not only structure, but exist within, the worldviews of group members (Gabora 2004; Gabora and Aerts 2009). This is not to say that there are no external indicators, linguistic formulations of, or pointers to cultural ideas; these are ubiquitous in material culture, normative rules, and social institutions. Such indicators have significance, however, only to the extent that individuals identify the external indicator with an understood meaning[8]. Thus, a major focus of child rearing across cultures is the forging of links in the child's mind between external indicators of cultural meaning, and internal experiences that substantiate them (e.g., Quinn 2006).

Although manifested across material and social vehicles, cultural idea systems exist primarily in the worldviews of group members, while simultaneously structuring these worldviews by providing frameworks for assimilation of, and accommodation to, experience (Gabora 2004, 2006). The idea systems, external indicators, and individual worldviews coevolve as the worldviews respond to private and social experience. Thus, we suggest that what evolves in cultural evolution, beyond changes in surface elements of culture, are worldviews (Gabora 2004, 2013, Gabora & Aerts, 2009, Gabora & Steel, 2017; Gabora & Smith, *in press*; Lane et al. 2009).

Worldviews lack the algorithmic structure necessary for their evolution to be modeled in analogy to the genetic case in which environmental selection on phenotypes results in changes in the frequency distribution of an underlying genotype. They do, however, possess the structure necessary to evolve through a more primitive, non-Darwinian form of evolution called *communal exchange* (Vetsigian, *et al* 2006). Each entity in a communal exchange process is self-organizing—a stable global organization emerges through interactions amongst its parts (Prigogine and Nicolis 1977). It is also autopoietic—it is subject to top down dynamic constraints that both maintain and reproduce its organizational structure (Maturana and Varela 1980). Like entities that evolve through variation-selection, entities that evolve through communal exchange produce self-variants; e.g., as is posited was the case with the early evolution of protocells prior to the appearance of RNA assembly codes (Farmer, *et al* 1986; Kaufman, 1993; Hordijk, 2013). Without a self-assembly code, the fidelity of this primitive replication in cultural evolution is lower than that of replication using a self-assembly code, but it is sufficient for cumulative, open-ended evolution.

Worldviews are resilient in that they self-organize in response to perturbations (such as inconsistencies or perceived threats to self-image), and they can be transmitted through social learning. Like an organism, a worldview is self-mending to the extent that people are inclined to explore possibilities and revise interpretations to establish or restore consistency (Osgood and Tannenbaum 1955; Greenwald *et al* 2002; Gabora and Merrifield 2012). A worldview is also self-regenerating: an adult shares knowledge and attitudes with children (and other adults), thereby influencing the conceptual closure processes by which worldviews form and transform. As they develop, children expose aspects of what were previously adult worldviews to different experiences, thereby weaving unique internal models of the relationship between self and world.

While we argue that worldviews evolve through mechanisms of communal exchange that operate at both the level of the individual worldview and the level of culture, we do not suggest that all changes in worldviews produce corresponding changes in cultural idea systems, e.g., the worldview of a person who migrates from a rural community to a large urban center may change substantially while the cultural idea systems that structure their understanding and interpretation of behavior might remain virtually unchanged.

**Cultural Idea Systems and Group Level Traits**

Cultural idea systems do not arise from the group-level synthesis of individual traits, either behavioral or genetic. Rather, they provide top down frameworks enabling individuals to interact in culturally appropriate ways, and individuals gain benefits only by behaving in accord with these prescribed forms. They establish the way that enculturated individuals interpret the experiences of everyday life, and their beliefs about how they ought to behave as participants in a jointly shared and objectified cultural reality (Hardin and Higgins 1996; Jost, *et al* 2007) if they are to remain members of the social collective and gain the corresponding fitness benefits.

Everett (2008) gives a good example of how a cultural idea system can structure social behavior and worldviews in his study of the Piraha, a tribe of about 300 to 400 people who live on the Maici River in the Brazilian province of Amazona. The Piraha language is perhaps the simplest language known and, among other unusual characteristics, it appears to lack recursion, subordinate clauses, and sense of time (although this has been challenged, e.g., see Reich 2012). The Piraha have no creation myths, and their cultural memory extends at best only slightly longer than a single generation. The explanation offered is that there is a primary cultural idea system that includes the injunction that one does not speak of what is not a matter of personal experience. This disallows complex, abstract thought and connections to the past, and constrains the language to be essentially a language of the present. This cultural idea system dominates all aspects of Piraha life so that selecting any particular manifestation of it as *the* associated group level trait is like trying to select a particular tree as representative of an entire forest.

Most researchers, though, use a bottom up approach focused on individual traits in order to attempt to understand social organization. Smaldino (2014, 2016) attempts to depart from this by making an analogy to the genotype-phenotype distinction, defining group level traits as "the phenotypic effects of social organization." In his view, just as selection acting on the biological phenotype leads to changes in gene frequencies, selection acting on emergent group phenotypes (the performance of groups in action) leads, he asserts, to changes in a group's organizational structure. The focus, however, remains on individual traits that allow effective role performance.

In addition, Smaldino's  discussion of cultural evolution using an analogy to biological evolution based on variation and selection is at best partial, and potentially misleading. The genome of an organism constitutes a self-assembly code that generates the phenotype while remaining shielded from direct external influence. Social groups, cultural idea systems, and worldviews lack self-assembly codes and their organizational structure is not shielded from direct selective action. Cooperative performance that depends on specialized roles introduces the relational system governing group organization as an additional factor that is independent of the individual traits of group members. Depending on how they are organized as a group, five young humans may constitute a variety of entities: a rock band; a basketball team; or, a criminal gang, for example, and the same five individuals could, conceivably, be all three. The organization is not found in the individual group members, nor does it emerge from their individual or group behavior. Rather, the organizational structure of group action is a socially given construct that defines the nature of the activity and does not exist outside of the social context. It is an institutional fact (Searle 2010; Smit *et al* 2011) that determines behavior and creates the behavioral result[9].

Cultural idea systems are distinct from the institutions, organizational systems, and norms that are their social manifestations. While these institutions and norms may potentially be considered as group-level traits, changes in a cultural idea system may not be indicated by corresponding changes in the associated institutions and norms, and vice versa. Because cultural idea systems are distributed across many aspects of group culture, determining associated group-level traits is problematic.

The distinction between a cultural idea system and its associated institutions and norms is illustrated by marriage (see Chit Hlaing and Read 2016). The *cultural idea system of marriage* is embedded in virtually every aspect of a society and, superficially, may be identified with formal institutions and customs. The *institution of marriage* in a society involves constraints such as who can marry, dowry or bride price, negotiated family alliances, celebratory rituals, civil or religious performances, sentimental expressions, and many other cultural elements. That there is a difference between the idea system of marriage and the institution of marriage is shown by the recent debate over same sex marriage in the United States (Read 2017). The terms of this debate show that it has not simply been a matter of extending the institutional character of marriage to same sex couples. Many opponents of same sex marriage were willing to allow equivalent institutional rights to be assigned to same sex partners, but same sex partners seeking to marry wanted something more than just institutional recognition of rights and privileges, while those opposing same sex marriage believed they were defending something sacred. What was at issue was that the *idea* of same sex marriage involved a change in the underlying cultural idea system, in particular, as it related to prescribed gender roles within a marriage, even as the institutional structures remained essentially unchanged.

**Kinship as a Cultural Idea System**

Kinship is the basic cultural idea system for assimilating a child into a social collective, and the functionality of kinship for a child stems from the collective identity that is provided by virtue of her or his kinship relations (Leaf and Read 2012). In societies such as hunter-gatherers, kinship relations link group members into a coherent whole, and provide the vocabulary for group members to communicate with each other regarding the interplay of collectively recognized kinship relations in everyday behavior (Read 2012a; Bird-David 2017).

While any discussion of early human behavior will be speculative, we believe that the origin of kinship systems, with the concomitant development of socially structured identities that become objects of reflective thought, played a pivotal role in a major change that occurred during the Middle to Upper Paleolithic in *Homo sapiens*—the uncoupling of social systems from space and time dependency (Gamble 2010; Leaf and Read 2012). Like present-day non-human primate social systems, earlier ancestral social systems for *Homo sapiens* would have been space dependent, with the maximal social group being composed of a collection of individuals living together. Time dependency is seen today in the fission-fusion form of chimpanzee social organization composed of small social units of males that are stable only on a time scale of hours to days (see references in Read 2012a).

Moffett (2013) makes a (non-exclusive) distinction between societies in which individuals are aware of other group members through direct recognition ("individual recognition societies") and those in which recognition is based on identifiable markings distinguishing those who are group members from those who are not ("anonymous societies"). Having distinctive group recognition signals allows group size to grow beyond the limits imposed by the necessities of individual recognition (e.g., Dunbar 1992). In this picture, initial boundaries of Paleolithic anonymous societies would have been established through recognizable group markings, beginning with distinct forms of non-linguistic vocalization. Without formal structuring, however, these markings would tend to drift, leading to division arising within a group as a population increased in size.

The space and time dependency of this form of social organization might be less than that of individual recognition societies, but identity signals for groups extended over wide spatial areas would drift, and the long-term temporal stability of group markings would be fragile. The best that could be expected would be a fission-fusion form of social organization in which a group might grow in size and spatial extent over one or two generations only to then collapse back to an aggregate of smaller groups. While early hominin societies likely developed group identification signals and markings, allowing within group cooperative behavior to some extent, these societies would have been unstable, resulting in social groups splitting as drift in group identifying marks took place and disputes arose without means for formal resolution (Moffett, 2013).

In the Upper Paleolithic, however, space and time dependent forms of social organization changed to something like the band-level form found in present day hunter-gatherer societies (Read 2012a; Read and van der Leeuiw 2017). Band-level social organization typically consists of a number of stable, spatially distributed residence groups, each built around family-level social units, with extensive integration across space and through time by the movement of families among residence groups.

Weissner (1998) suggests that the transition to band level organization was facilitated by an evolved psychological susceptibility to indoctrination. This, in addition to well-established social conformity biases (e.g., Asch 1955) allowed homogenization of behavior and understanding across local residence groups. This allowed the formation of (at least partially) anonymous societies, but more is required to stabilize these societies against spatial and temporal drift. We suggest that the essential element was formation of formal kinship systems, developed by our ancestors during the Upper Paleolithic (Leaf and Read 2012: Chapter 3). Kinship systems produce a stable space and time independent form of social relations for a society as a whole, hence are the organizational key to band-level social structures (Read 2015a). The appearance of the hunter-gatherer type of social systems in the Upper Paleolithic would not have been possible without the development of conceptually formulated systems of kin relations (Read and van der Leeuw 2015), eventually encapsulated in kinship terminologies that made it possible for participants in social interaction to recognize their respective positions in a shared, culturally constructed system of kinship relations (Leaf and Read 2012; Read 2015a,b).

Kinship systems include prescriptions to the effect that those who are kin to one another should be mutually supportive, as expressed in Forte's (1969) Axiom of Prescriptive Altruism. Prescription, of course, need not translate into behavior, and in reality kin can be selfish, uncooperative, and mean spirited. What is critical is that those breaching prescriptions can be called to account by their kin.

In hunter-gatherer societies, those who are not kin are strangers, and may be treated as enemies (Read 2012a). At least before European contact led to major changes in their traditional ways of life, altruistic behavior in hunter-gatherer bands, especially with regard to strangers from beyond the culturally recognized group, was uncommon (Marlowe 2010). Referring to the !Kung San, Weissner (2009:133) notes that "There is little evidence… that it is a part of human psychology [for them] to be willing to engage in altruism… in a social and cultural vacuum. When the faces and forces of culturally defined institutions are reintroduced, sharing and giving resume."

Among the Hadza, sharing occurs because of kinship obligations or relations, such as a man providing meat for his his children or his wife's mother (Marlow 2010: 170) or because of the demands of others when they see someone with meat (Marlow 2010: 251). Rather than sharing meat altruistically, Hadza men will hide meat so as to avoid having to meet sharing obligations (Marlow 2010: 238 – 239). In general, as discussed in Read (2012b), in hunter-gatherer societies the members of a residence group own food resources in the wild collectively and it is the members of the residence group that have the right of access to those resources. Low risk resources (food sources that are abundant, small in size, predictable with regard to their occurrence, and only requiring skills common to all adults to procure them) are culturally transformed into individual ownership when procured by an individual. High risk resources (food resources that are large, variable and unpredictable, and requiring special skills to procure effectively) are not transformed culturally into individual ownership by the actions of a hunter. Instead, collective ownership is transformed into individual ownership by cultural rules for the distribution of the hunted animal. Altruism, as an analytical concept, only applies to items that are individually owned; outside of the family what is individually owned is not distributed altruistically.

The concept of a system of kinship relations expressed symbolically through kin terms, with a computational logic that enables kinship relations to be worked out through those kin terms (Read 2007), provides the foundation for the ties binding individual worldviews into a collective cultural worldview that can act as a seed for the stable transmission of cultural idea systems across generations (Read 2012a). This provided a strong evolutionary advantage for both individuals and groups: "The evolution of socially defined kinship… permitted the construction of broad social security networks for risk reduction by granting access to human and natural resources lying outside the [immediate] group" (Weissner 1998: 134).

Although the appearance of kinship systems seems to presuppose a pre-existing degree of cooperation, we do not believe this gives rise to a chicken or egg problem. From the time of the divergence of the ancestors of *Homo* and of *Pan*, the evolutionary path to *Homo sapiens* exhibits several interconnected threads, including changes in morphology, technology, diet, encephalization, changes in social behavior, division of labor, male-female relations, cognitive abilities, and so on. In particular, greater cognitive capacities enabled a cultural means to circumvent the cognitive bottleneck faced by the African great apes arising from individualized behavior coupled with social organization based on face-to-face interaction that, if left unchecked, would increase social complexity exponentially (Read 2012a). The critical, culturally based social difference that distinguishes *Homo sapiens* from other primates is the escape from this bottleneck that was provided by a shift from social organization based on face-to-face interactions to a culturally grounded, relationally based form of social organization (Read 2012a). Moffett (2013) sees this in terms of a transition from individual recognition societies (face-to-face) to anonymous societies. However, as we have indicated, more than the establishment of group recognition signals is required; namely, the working out of culturally transmitted systems of kinship relations.

Cognitively, this required the capacity for *self-triggered recursive thought*, which enables manipulation of raw experiences such that they cohere with and relate to one another, often achieved through the development of stories and narratives (Gabora, 1999, 2000; Gabora & Steel 2017). It additionally requires *contextual focus:* the ability to shift between different modes of thought, a convergent mode conducive to mental operations involving relationships of causation, and a divergent mode conducive to mental operations involving correlation, or similarity (DiPaola and Gabora 2009; Gabora 2003, 2010). These two abilities enabled them to not only structure goal directed sequences of thought, which is within the cognitive capacity of other primates, but to grasp abstract relations and then manipulate and apply them recursively (Gabora and Kitto 2013; Gabora and Smith *in press*; Read 2012a). For example, the ability to abstract the concept of a mother-child relation from observation of incidents of mothering behavior, and then the ability to compose this relation to arrive at the mother of a mother, or grandmother concept. This provides the basis for working out other genealogical relations and, eventually, bootstrapping from genealogical terms to formal kinship systems. Group boundaries are then determined by those who can recognize that they are kin to each other (Read 2012a). The ability of two individuals to establish their kin relationship to each other was stabilized through the development of a computational system of kin relations we refer to as a kinship terminology (Read 2007). Working out a computational system of kin relations was a remarkable intellectual *tour de force* that likely occurred during the Upper Paleolithic (Leaf and Read 2012). The functionality of this evolved, socially constructed system of kin relations could not have been realized without transmission through enculturation, which also establishes coordination and shared understanding among group members (Read 2010, 2012a: 165-167).

**Emotions, Feelings, and Identity**

We have argued that cultural idea systems, and the individual worldviews that these idea systems both structure and are carried by, coevolve primarily through mechanisms of communal exchange. These idea systems incorporate prescriptions for cooperative behavior, and the individual identities established through these idea systems thus express expectations of cooperation with those culturally recognized as kin, and self-sacrifice for the benefit of the group. In this section we ground this argument in individual psychology and biology.

Humans congregate in marked groups and make personal sacrifices for the benefit of the group even if it is not composed of close genetic kin, and even if no personal benefit, either direct or indirect, accrues. In culture-gene theories this is explained via group selection for social emotions, evolved as individual genetic adaptations that are favored because they supported non-biological kin cooperation in small human groups under conditions where such cooperation confers substantial group benefits (Bowles and Gintis 2011; Moffett 2013; Tomasello 2014). What is lacking is an explanation of why these emotions exert such powerful influence on behavior. One assumption might be that the experience of feelings such as guilt or shame is unpleasant, hence people tend to avoid behavior resulting in such feelings. While this idea carries some weight, shame and guilt are easily quelled through indignation and self-justification so additional assumptions are required; for example, that they act preemptively to deter behavior that is likely to incur social censure (Bowles and Gintis 2011).

In contrast, we propose that the psychological power of the social emotions is found in the human capacity for (and vulnerability to) identification with ideas, beliefs, symbolic cues, and other group markings; that is, indoctrination (Weissner 1998). Group membership and the associated group markings, norms, and ideals, are interjected as part of an enculturated identity (Atran 2016; Abrams *et al* 1990; Heine 2001; Burke and Stets 1999; Stets and Burke 2000; Ellemers, *et al* 2002; Hornsey 2008; Swann Jr., *et al* 2014; Weissner 1998). Through this process a linkage is formed between social identity and survival related circuits in the brain that are recruited for defense of both social identity and the groups upon which this identity depends. This linkage arises through a two-way interaction of emotions and feelings[10].

An *emotion* is a complex of physiological responses to survival related stimuli. Associated with this is the qualitative *feeling* that arises in the mind as a result of the emotional response (Damasio 2010). While emotions are directly connected to biological survival instincts, the associated feelings are culture-laden. Cultural idea systems provide interpretations for feelings and prescribe culturally acceptable channels for their expression (Mesquita and Walker 2003; Tsai *et al* 2007; Koopmann-Holm and Matsumoto 2011; De Leersnyder, *et al* 2013; Gervais and Fessler 2017). These same cultural idea systems contribute to social identity through valorization of group membership and role performance, often grounded in painful past shared experiences (Whitehouse *et al* 2017) that lead to identity fusion between the individual and the group (Swann Jr., *et al*).

While some researchers link emotions to dedicated circuits in the brain evoked by salient stimuli, others relate them to more general survival circuits found in many species (Phillips *et al* 2003; LeDoux 2012; Pessoa and McMenamin 2016). *The essential point is* *that the neural basis for the emotions is tied directly to biological survival, while the corresponding feelings can be evoked through culturally loaded cues and, when a feeling is so evoked, it can produce an emotional response that reinforces the feeling in a positive feedback loop* (Damasio 2010). As a result, apparent threats to social identity, or to the groups in terms of which this identity is defined, can produce emotional responses grounded in powerful instincts of biological survival. We are thus able to address two issues faced by theories of cooperation: the problem of assortment, and the second-order free rider problem.

Mathematical and simulation models suggest that in order for a cooperative group to avoid invasion by defectors, cooperators must preferentially assort with other cooperators (Nowak and May 1992; Nowak and Sigmund 2004; Lehmann and Keller 2006; Jackson and Watson 2013). Further, they must punish non-cooperators and even those who fail to cooperate in punishing. The concept of *strong reciprocity* was introduced to describe this behavior (Fehr, *et al* 2002; Bowles and Gintis 2004) but, as with ultracooperation, strong reciprocity is uniquely human, and, arguably, could not have evolved through processes involving only individual inclusive fitness (Fehr, *et al* 2002; Stephens 2005; Read 2012b).

The *assortment problem* is to determine how cooperators can carry out the preferential assortment required to exclude free riders and the second order free rider problem arises from the question of whether or not individuals who agree to participate in punishment of defectors will actually do so if the need arises. We address both assortment and free riding in terms of the culturally determined social identities taken on by individual group members. These identities are interlinked through shared cultural idea systems that provide a framework for preferential assortment, prescribe expected behavior in social interactions, and cue defensive emotional reactions when faced with defection.

To the extent that a person identifies with a group or a social role, they will experience feelings of guilt and/or shame at failing to live in accord with that identification, and will feel validated, enhanced, and justified at successful adherence to their social self-image. Expression of such feelings, or exhibition of physiological signs of the associated emotions, is a signal to others that an individual adheres to shared norms of cooperation, hence is a suitable partner for cooperative behavior. Jablonka *et al* (2012), for instance, suggest that blushing evolved as a social signal of emotional response indicating, in context, the presence of a particular feeling and the corresponding emotion.

To address the second order free rider problem, Bowles and Gintis (2011) argue that punishment involves an innate impulse of retribution, but do not account for how such an innate impulse arises. Our theory provides a detailed account of how this can happen. Perceived deviations from expected behavior can elicit feelings that trigger threat responses, evoking emotions that are felt as indignation and anger (Ellemers, *et al* 2013; van der Toorn, *et al* 2015). This can serve to intensify the initial feelings in a positive feedback cycle that results in punishing deviant behavior. In this way, maintaining social identity, and the social group that supports it, is directly connected to the imperatives of biological survival. This manifests in in-group/out-group conflict as well when perceived threats to group identification marks trigger defensive responses: "the perception of threats to symbols is a common issue in aggression toward other societies" (Moffett 2013: 249).

There is also a degree of internal control over the temptation to defect. Going against a social or cultural expectation or norm may be self-perceived as a justifiable personal attempt to gain benefit, carrying an associated danger of punishment, but it may also be experienced as a denial of identity, requiring defensive justification. Disruption of the patterns of relationships inherent in what is experienced as "one's self," presents a threat to that self (Branscombe *et al* 1993; Ellemers, *et al* 2002; Ellemers 2012; Tritt, *et al* 2012). Similarly, the perception that another group member is violating an expected pattern of social behavior cues feelings/emotions motivating reactions of indignation, condemnation, and an impulse to punish (Ellemers, *et al* 2013; Van der Toorn, *et al* 2015). In sum, punishment of defectors does not arise from previous agreement, but as an immediate felt emotional impulse, triggered as a response to identity threat, and enacted through forms prescribed by a cultural idea system. Hence, no agreement to punish is required and yet punishing behavior will occur when faced with individuals exhibiting deviance from expected behavior by those exhibiting marks or signals of group membership.

The problem of second-order free riders only arises if the question is how a group of inherent cooperators can resist cheaters. Starting from the premise that, biologically, individuals are all cheaters, the problem is not one of punishing cheaters; it is how to build a cultural idea system and enculture individual worldviews with that system such that the cultural meanings and expectations of the associated social identities exclude cheating. The cultural idea system overrides biological impulses that say, "be a cheater," because the cost of cheating is high. To put it simply, being recognized as a cheater defines an individual as an untrustworthy kinsman in a context in which survival depends on having good relations with one's kin (Read 2012b).

**Summary and Discussion**

We propose that human cooperation should be viewed as a component of how individuals understand who they are by virtue of the cultural context within which they are embedded, which expresses cooperative behavior as part of a culturally defined identity. The cultural identity of a !Kung San hunter, for example, says that he cannot act as an individual with individual interests, as expected from a biological model of behavior; rather, as a cultural individual he is expected, *and expects*, to act cooperatively.

Through the evolution of the capacity to weave individually and culturally acquired knowledge into a self-organized, loosely integrated worldview (Gabora, 1999; Gabora & Aerts, 2009; Gabora & Steel, 2017; Gabora & Smith, in press), hominids became capable of reflective self-consciousness. This permitted the conceptual abstraction of self and group identities, eventually enabling the development of kinship systems as a stable basis for cooperative behavior with those identified as cultural kin.

Kinship and other cultural idea systems exert top down control on group members through patterns of behavioral expectations and obligations that transform basic biological response impulses into socially prescribed behavior (Markus and Kitayama 1994; Weissner 1998; Ochsner and Gross 2005; Barrett *et al* 2007; Cikara, *et al* 2011). Multi-level selection favors cultural idea systems that provide channels of behavioral expression for biological impulses such that coordinated group cooperative living becomes possible. Beginning in early infancy, enculturation, as a process of indoctrination (Weissner 1998), establishes psychological barriers against deviance that exist with sufficient strength that in some cases even a deviant thought is felt as a threat, and suppressed.

This multi-level organization, in which cultural idea systems structure individual worldviews that recursively act as hosts for, and transmitters of, these idea systems, is predicated on the evolved capacity for abstraction of group identities from aggregates of similar entities, and the concomitant capacity for reflective thought about these identities. It is this that enables humans to conceptualize themselves as individuals who are also identified as group members, while simultaneously tying their individual identity to top down behavioral prescriptions associated with that membership.

Psychologically, the linkage of feelings and emotions connects social identity to the individual biological self, which then leads to recruitment of survival related neural circuits for defense of the social self and the groups on which this self relies. Biological defensive reactions such as fight, flight, or freeze are transformed into multiple forms of culturally mediated response behavior, including indignation at, and an impulse to punish, those perceived to violate one's cultural identity expressed through cultural idea systems and worldviews.

What we have presented here is a synthesis of research from multiple disciplines into a theoretical framework that can explain human cooperation. Directions of research highlighted by our theory are the need to: (1) Determine and describe the relations between the cognitive, emotional, social, and psychosomatic aspects of personal and social identity; (2) Determine the mechanisms by which external inputs and cues evoke responses energized by survival related neural circuits; (3) Clarify the neural circuits underpinning behavioral responses to both positive and negative social input; (4) Construct testable models of multilevel selection acting on cultural idea systems and worldviews (c.f.,  Read 1987); (5) Consider whether there is a deeper aspect of identity that appears in the direct recognition of an other as another like myself, beyond social and cultural conditioning (e.g., Sahlins 2013).

Extensive research is being carried out in many fields seeking to elucidate cultural and psychological aspects of ultrasociality, social identity, emotion, how these are connected, the underlying neural circuits, and how these factors interact in behavior (e.g., Phillips *et al* 2003; De Dreu *et al* 2010; Cikara, *et al* 2011; Dawes *et al* 2012; Jablonka, *et al* 2012; Cikara and Van Bavel 2014; Jensen, *et al* 2014; Nummenmaa *et al* 2014; Özkarar-Gradwhol *et al* 2014; Chang, *et al* 2016). To date, however, this work has not been integrated into a unified understanding of human behavior. We hope the theory of human cooperation provided here will make a fruitful contribution.

**Acknowledgments**

This work has been partially supported by NSERC grant 62R0653 to Liane Gabora. The theory outlined in this paper was presented in outline form at the 2015 (Phoenix) and 2016 (Amsterdam) Complex Systems Conferences, in a seminar at Santa Fe Institute in April 2016, and at the inaugural Conference on Cultural Evolution (Jena) in September 2017. We thank Samantha Thomson for assistance in preparation of the manuscript and several anonymous referees for their detailed comments on earlier versions of this manuscript.

**End Notes:**

---

[1] This can be multi-layered as indicated by the traditional Bedouin saying: "I against my brothers; my brothers and I against our cousins; I, my brothers, and our cousins against the world."

[2] We realize that the coherence and unity provided by cultural idea systems may only be apparent, that there may well be internal contradictions. The point is that for individuals enculturated within a cultural idea system, it does appear coherent and unified. When contradictions do appear the reaction is to produce rationalizations that explain them away (e.g., Mercier and Sperber 2017).

[3] The word syntonization well captures our meaning, it refers to (a) the establishment of communication between distinct electronic circuits through bringing them to the same frequency; and (b) to the tuning of voices in a choir to produce a harmonious performance.

[4] When this top down behavioral control breaks down, explosions of violent response can occur and the frequency of such occurrences gives a measure of the degree of stability of social control.

[5] This is far more threatening in small hunter-gatherer bands where the possibility of ostracism exerts powerful social control. For example, see the story of Cehpu related in Turnbull (1961). The threat remains in all societies, however, and can exert a powerful conformist influence (Lessing 1986).

[6] We bear in mind, however, that "it isn't known whether primates form societies by primates recognizing and bonding to the group as a whole… rather than bonding to other members individually." (Moffett 2013: 227) so further research in this area is required.

[7] Our approach is closely related to recent work of McGeer (e.g., 2015) and Zawidski (2013) and we thank an anonymous referee for bringing this to our attention.

[8] There is an analogy between cultural idea systems and the way that mandalas are used in Tibetan Buddhism. In appearance, a mandala is structured as a collection of images surrounding a central image (usually a specific deity). The central image itself contains a number of items such as weapons, bells, nooses, and so on. Superficially, a follower of the Tibetan system seems to use a mandala simply as part of a meditative exercise. In fact, the meditator will have spent many hours of preparatory meditation on the details of the various elements of the mandala, developing a direct understanding of their meaning in the Tibetan system. The mandala acts as a synthetic image that correlates all of its various elements into a coherent gestalt that conveys information descriptive of a psychological state that the meditator seeks to experience in greater depth (Leidy and Thurman 1997). In the same way, an enculturated individual sees multiple elements of cultural idea systems in their everyday environment, which act both to

reinforce enculturation and provide a framework of understanding and familiarity. The entire complex of cultural ideas could well be termed a "cultural mandala."

[9]Smit and his collaborators take a very different view of "institutional facts" than does Searle, but both agree on the social nature of such "facts."

[10] There is lack of agreement in the emotion literature as to a label for what we are calling feelings and emotions. Gervais and Fessler (2017), for example, use the terms "basic affective system" (emotions) and "folk affective concepts" (feelings). Our usage follows Damasio (2010).